\documentstyle[12pt,fleqn]{article}
\author{L.Didukh, V.Hankevych$^{\dagger}$ \\
{\small \em Ternopil State Technical University, Department of Physics,}\\ 
{\small \em 56 Rus'ka Str., Ternopil UA--282001, Ukraine; Tel.:+380352251946,}
\\ 
{\small \em Fax: +380352254983; E-mail: didukh@tu.edu.te.ua} \\
{\small \em $^{\dagger}$Email: vaha@tu.edu.te.ua}}

\date{}
\title{Temperature-induced metal-insulator transition in a narrow-band
model with non-equivalent Hubbard subbands at half-filling}

\begin{document}
\maketitle

\begin{abstract}
In the present paper the temperature-induced metal-insulator transition 
is studied in a generalized Hubbard model with correlated hopping using
recently obtained expression for energy gap. 
The dependence of energy gap on concentration of doubly occupancy  
leads to increasing energy gap width with increase of temperature. 
Thus narrow-band system can undergo transition from a metallic state 
to an insulating state with the increase of temperature.

For some values of intra-atomic Coulomb repulsion $U$ and $w_0$ ($w_0$ is 
half-bandwidth without taking into 
account of correlated hopping of electrons) we find the values of 
temperature when narrow-band material undergoes transition from a metallic 
state to an insulating state. We show that at given $U/w_0$ metal-insulator 
transition in model with non-equivalent Hubbard sub-bands can occur at 
smaller temperature than in the Hubbard model. It testifies on 
the fact that taking into account
of correlated hopping is important for a consideration of  
metal-insulator transition problem.

The obtained results are applied to the interpretation of the experimental 
data.
\end{abstract}

\section{Introduction}

It is known that the electron-hole symmetry is peculiar to the Hubbard 
model~\cite{1}.
One of manifestations of this symmetry is the equivalence of the lower and 
upper Hubbard bands. It is the result of an equality of hopping integrals
describing both ``translational'' hoping of holes and doubly occupied sites
(doublons) and the processes of their paired creation and destruction. 
Equality of noted hopping integrals is caused by neglecting the matrix 
elements of electron-electron interaction
\begin{eqnarray}
J(ikjk)=\int \int \phi^*({\bf r}-{\bf R_i})\phi({\bf r}-{\bf R_j})
{e^2\over |{\bf r}-{\bf r'}|}|\phi({\bf r'}-{\bf R_k})|^2{\bf drdr'},
\end{eqnarray}
in the general Hamiltonian (the matrix elements~(1) describe hopping of 
electrons between $i$ and $j$ lattice sites; $\phi$-function is the Wannier 
function). 

However, theoretical analysis, on the one hand, and available experimental 
data, on the other hand, point out the fact that the Hubbard model generalization
by taking into account correlated hopping~(1) is of principle 
necessary~\cite{2} -- \cite{5}. In such model hopping integrals
describing ``translational'' hopping of
holes and doublons are different. These hopping integrals also differ
from the hopping integral which is connected with the processes of paired 
creation and destruction of holes and doublons. In consequence of that 
the lower and upper Hubbard bands are non-equivalent (non-symmetric). 
In recent years similar models have been studied 
intensively~\cite{7} -- \cite{11}.
 
Important puzzle arising in an investigation of these models is 
metal-insulator transition problem which is one of the most essential in
narrow-band \mbox{physics~\cite{12} -- \cite{14}.} In this connection special 
interest is challenged by the observable metal-to-insulator transitions in 
some narrow-band materials with the increase of temperature (see, for
example~\cite{14} -- \cite{18}).

On the basis of an approach proposed in the papers~\cite{3,19}
we have studied metal-insulator transition in a model of narrow-band material
with non-equivalent Hubbard subbands (so-called ``non-symmetric Hubbard 
model'') at half-filling and zero temperature in
the paper~\cite{20}. The present paper is devoted to a further study of
metal-insulator transition in a model with non-equivalent Hubbard subbands,
in particular, an investigation of temperature-induced metal-to-insulator
transition.

\section{Results}

We start from the following natural generalization of the Hubbard model~\cite{1} 
at half-filling including correlated hopping~(1) \cite{2,3,20}
\begin{eqnarray}
H=&&-\mu \sum_{i\sigma}a_{i\sigma}^{+}a_{i\sigma}+
(t_0+T_1){\sum \limits_{ij\sigma}}'a_{i\sigma}^{+}a_{j\sigma}+
T_2{\sum \limits_{ij\sigma}}' \left(a_{i\sigma}^{+}a_{j\sigma}n_{i\bar \sigma}+h.c.\right)
\nonumber\\
&&+U \sum_{i}n_{i\uparrow}n_{i\downarrow},
\end{eqnarray}
where $\mu$ is the chemical potential, $a_{i\sigma}^{+}, 
(a_{i\sigma})$ is the creation (destruction) operator of an electron of spin 
$\sigma$ ($\sigma=\uparrow, \downarrow$) on $i$-site  
(${\bar \sigma}$ denotes spin projection which is opposite to $\sigma$),
$n_{i\sigma}=a_{i\sigma}^{+}a_{i\sigma}$ is the number operator of electrons
of spin $\sigma$ on $i$-site,  $U$ is the intra-atomic Coulomb repulsion,
$t_0,\ T_1,\ T_2$ are the integrals 
of electron hopping between nearest neighbors,
\begin{eqnarray*}
T_1=\sum_{\stackrel{k\neq{i}}{k\neq{j}}}J(ikjk), \quad
T_2=J(iiij);
\end{eqnarray*}
the primes at the sums in Hamiltonian~(2) signify that $i\neq j$.

Using  a generalized mean-field approximation~\cite{3,19} in Green function 
method we obtain for a paramagnetic state the 
single-particle energy spectrum as~\cite{20}
\begin{eqnarray}
&&E_{1,2}({\bf k})=-\mu+{(1-2d)(t_{\bf k}+\tilde{t}_{\bf k})+U\over 2}\mp 
{1\over 2}F_{\bf k},
\\
&&F_{\bf k}=\sqrt{\left[B(t_{\bf k}-\tilde{t}_{\bf k})-U\right]^2+
(4dt'_{\bf k})^2},\quad
B=1-2d+4d^2,
\end{eqnarray}
where ${\bf k}$ is the wave vector, $E_1({\bf k}),\ (E_2({\bf k}))$ are the 
energies of electrons within the lower (upper) Hubbard band, $d$ is the 
concentration of polar states (holes or doublons), $t_{\bf k},\ 
\tilde{t}_{\bf k},\ t'_{\bf k}$ are the Fourier transforms of respective
hopping integrals $t=t_0+T_1,\ \tilde{t}=t+2T_2,\ t'=t+T_2$; $t$ and 
$\tilde{t}$ are terms describing hopping of quasiparticles within the lower 
and upper Hubbard bands (hopping of holes and doublons) respectively, $t'$
describes quasiparticle hopping between hole and doublon bands (the processes 
of paired creation and destruction of holes and doublons).

Note that single-particle energy spectrum~(3) gives the exact atomic and
band limits: if $U=0$ and $t_{\bf k}=\tilde{t}_{\bf k}=t'_{\bf k}=
t_0({\bf k})$ (neglecting correlated hopping) then 
$E_{1,2}({\bf k})$ take the band form, if 
$t_{\bf k}=\tilde{t}_{\bf k}=t'_{\bf k}\rightarrow 0$ then we obtain exact
atomic limit.

With the help of single-particle energy spectrum~(3) we find the expression
to calculate the energy gap width (difference of energies between bottom of 
the upper and top of the lower Hubbard bands):
\begin{eqnarray}
&&\Delta E=-(1-2d)(w+\tilde{w})+{1\over 2}(Q_1+Q_2),
\\
&&Q_1=\sqrt{\left[ B(w-\tilde{w})-U\right]^2+(4dzt')^2},\\ 
&&Q_2=\sqrt{\left[ B(w-\tilde{w})+U\right]^2+(4dzt')^2}, 
\end{eqnarray}
where $w$ and $\tilde{w}$ are halfwidths of the lower (hole) and upper 
(doublon) Hubbard bands: $w=z|t|,\ \tilde{w}=z|\tilde{t}|$ ($z$ is the 
number of nearest neighbors to a site).

At given $w,\ \tilde{w},\ t',\ U$ and change of $d$ (which can be caused by
external influences, in particular, by temperature) energy gap~(5) vanishes
when the condition $d\leq d_0$ is satisfied where $d_0$ is the root of the
equation
\begin{eqnarray}
-(1-2d)(w+\tilde{w})+{1\over 2}{(Q_1+Q_2)}=0.
\end{eqnarray}
Then a metallic state ($\Delta E\leq 0$) is realized at $d\leq d_0$, an
insulating state ($\Delta E>0$) is realized at $d>d_0$.

If $t=\tilde{t}=t'=t_0$ (the Hubbard model) then energy gap~(5) takes the 
following form~\cite{19}:
\begin{eqnarray}
\Delta E=-2w(1-2d)+\sqrt{U^2+(4dw)^2},
\end{eqnarray}
and vanishes when the condition $d\leq d_0$ is satisfied where
\begin{eqnarray}
d_0={1-(U/2w)^2\over 4} \qquad (2w\geq U).
\end{eqnarray}

From Eq.~(8) we obtain the dependence of $d_0$ on $U/w$ ratio (Fig.~1). 
The parameters $\tau_1=T_1/|t_0|,\ \tau_2=T_2/|t_0|$ characterize
value of correlated hopping. From Fig.~1 one can see that value of $d_0$ 
depends on the parameters of correlated hopping $\tau_1,\ \tau_2$ 
(thus on $\tilde{w}/w$) weakly when $U/w$ is near zero. But with the increase 
of $U/w$ the value of $d_0$ begins to depend strongly on the parameters 
$\tau_1,\ \tau_2$ . Fig.~1 points out the extension of region in which 
narrow-band material is a metal with the increase of $\tilde{w}/w$
ratio (at given $U/w$).

The concentration of polar states (obtained with the help of the Green
function $\langle\langle a_{i\sigma}n_{i\bar{\sigma}}|a_{j\sigma}^{+}\rangle
\rangle$) is~\cite{20}
\begin{eqnarray}
d={1\over 4w}\int\limits_{-w}^{w}\left[{C_{\varepsilon}\over \exp{E_1(\varepsilon)
\over k_BT}+1}+{D_{\varepsilon}\over \exp{E_2(\varepsilon)\over k_BT}+1}
\right]d\varepsilon
\end{eqnarray}
with
\begin{eqnarray}
&&C_{\varepsilon}={1\over 2}-{U\over 2F_\varepsilon}-
{B\varepsilon\over 2F_\epsilon}\left({\tilde{t}\over t}-1\right),\\
&&D_{\varepsilon}={1\over 2}+{U\over 2F_\varepsilon}+
{B\varepsilon\over 2F_\varepsilon}\left({\tilde{t}\over t}-1\right),
\end{eqnarray} 
and chemical potential of narrow-band model with non-equivalent Hubbard
subbands is given by the equation
\begin{eqnarray}
\int\limits_{-w}^{w}\left[{1\over \exp{-E_2(\varepsilon)
\over k_BT}+1}-{1\over \exp{E_1(\varepsilon)\over k_BT}+1}
\right]d\varepsilon=0,
\end{eqnarray}
$E_{1,2}(\varepsilon),\ F_\varepsilon$ are obtained from respective 
formulae~(3), (4) for $E_{1,2}({\bf k}),\ F_{\bf k}$ by substitution of
$t_{\bf k}\rightarrow \varepsilon,\ \tilde{t}_{\bf k}\rightarrow (\tilde{t}/t)
\varepsilon,\ t'_{\bf k}\rightarrow (t'/t)\varepsilon$. Here we have used
the rectangular density of states
\begin{eqnarray}
{1\over N}\sum_{\bf k}\delta(E-t({\bf k}))={1\over 2w}\theta(w^2-E^2),
\end{eqnarray}
where $\theta(x)=1$ if $x>0,\ =0$ otherwise; $N$ is the number of lattice 
sites.

\section{Discussion}

The temperature dependence of chemical potential of a narrow-band model with
non-equivalent Hubbard subbands obtained from Eq.~(12) is plotted in Fig.~2.
One can see that in the considered model in the region of low and normal 
temperatures chemical potential is essentially dependent not
only on the parameters $w$ and $\tilde{w}$ and also on temperature (in 
contrast to the Hubbard model where $\mu=U/2$), and what is more with
the decrease of temperature chemical potential rapidly increases depending on
the paramaters of non-equivalence of Hubbard bands $\tau_1,\ \tau_2$. In 
high temperature region in the proposed model chemical
potential tends to $U/2$ with the increase of temperature; really, at 
$T\to \infty$ the probabilities of an electron finding within the lower and
upper Hubbard bands (independently of their bandwidths ratio) are equal. 

At given $U,\ w,\ \tilde{w},\ t'$ (constant exterior pressure) concentration
of polar states~(11) increases with the increase of temperature. It leads to
the fact that system can undergo transition from the state with 
$\Delta E\leq 0$ to the
state with $\Delta E>0$, i.e. metal-to-insulator transition can occur. 
In this case the results obtained in the Hubbard model and those obtained in 
non-symmetric Hubbard model can be essentially different (Fig.~3 illustrates
it). Let us take for example $U/w=0.9$. One can see that at $T=0$ the
energy gap width in both models is $\Delta E<0$ (a metallic state). With the
increase of temperature metal-to-insulator transition does not occur in the
Hubbard model, in non-symmetric model the values of parameters 
$\tau_1,\ \tau_2$ exist at which metal-to-insulator transition occurs. 

In case metal-to-insulator transition occurs in both models from Fig.~3 one
can see that at given values of $U/w$ in model with non-equivalent Hubbard 
subbands metal-to-insulator transition occurs at smaller temperature than in
the Hubbard model. So, for example, for $w_0=z|t_0|\approx\!1.05$~eV
(such bandwidth of NiS$_2$ was estimated in paper~\cite{16})
in considered model in a paramagnetic state metal-to-insulator transition 
occurs at $T\approx\!280$~K for $U/w_0=1.94$ and $\tau_1=\tau_2=0.01$ 
(observable the transition temperature of NiS$_2$ is $T\sim\!280$~K at 
$p\sim\!3$~MPa~\cite{17}). For the same value of $U/w_0$
metal-to-insulator transition occurs at $T\approx\!940$~K when 
$\tau_1=\tau_2=0$ (neglecting of correlated hopping, the Hubbard model) and 
at $T=\!0$~K when $\tau_1=\tau_2=0.015$. If $U/w_0=1.98$ then transition from a 
metallic state to an insulating state is realized at $T\approx\!290$~K for 
$\tau_1=\tau_2=0$; at $T=\!0$~K when $\tau_1=\tau_2=0.005$. Note that
at $U\sim 2w$ the temperatures of metal-to-insulator transiton found in
both models are essentially different; with a deviation from this ratio
the difference decreases.

The obtained temperature dependence of energy gap can explain observable
transition from the state of a paramagnetic  metal  to  the  paramagnetic
Mott-Hubbard insulator state in the (V$_{1-x}$Cr$_x$)$_2$O$_3$ compound~\cite{14}, 
\cite{15} in NiS$_2$~\cite{17} and in the NiS$_{2-x}$Se$_x$ system~\cite{17,18} with the increase of temperature.

In summary, metal-to-insulator transition observable in the materials with
narrow energy bands can be explained on the basis of the proposed approach at
realistic values of the parameters characterizing non-equivalence of the
Hubbard subbands. In this way series of results obtained from such 
consideration are essentially distinct from those obtained without taking
into account this non-equivalence (the Hubbard model). In particular, 
the temperature of metal-to-insulator transition is essentially smaller than in 
the Hubbard model (this fact agrees to the observable transition 
temperature), and chemical potential disctinguishes from the value $\mu=U/2$
(obtained in the Hubbard model) and is temperature dependent.

Figure~1: The dependence of $d_0$ on $U/w$ ratio: the upper curve corresponds
to $\tau_1=\tau_2=0$ (the Hubbard model); the middle curve -- $\tau_1=\tau_2=0.1$; 
the lower curve -- $\tau_1=\tau_2=0.2$. 
To the left of respective curve there is a metallic phase, to the right --
an insulating phase.

Figure~2: The temperature dependence ($\theta =k_BT$) of chemical 
potential $\mu$ for $U/2w=1$:
the upper curve corresponds to $\tau_1=\tau_2=0.3$; the lower curve
-- $\tau_1=\tau_2=0.1$; the straight line corresponds to values of 
chemical potential in the Hubbard model ($\tau_1=\tau_2=0$).

Figure~3: The dependence of energy gap on temperature at $U/w=0.9$. 
The upper curve corresponds to $\tau_1=\tau_2=0.2$,  
the middle curve -- $\tau_1=\tau_2=0.1$, the lower curve -- 
$\tau_1=\tau_2=0$ (the Hubbard model).  

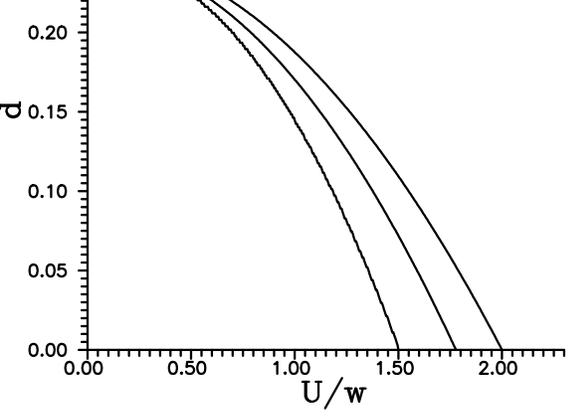
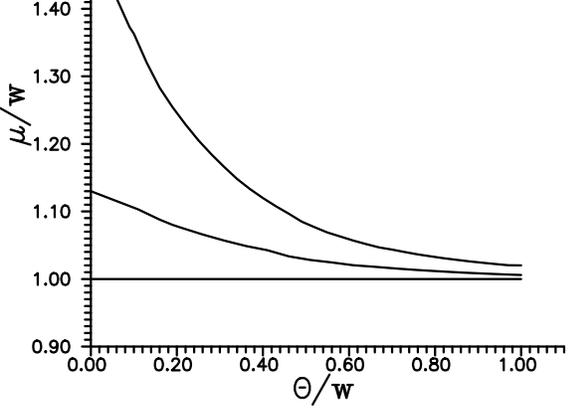
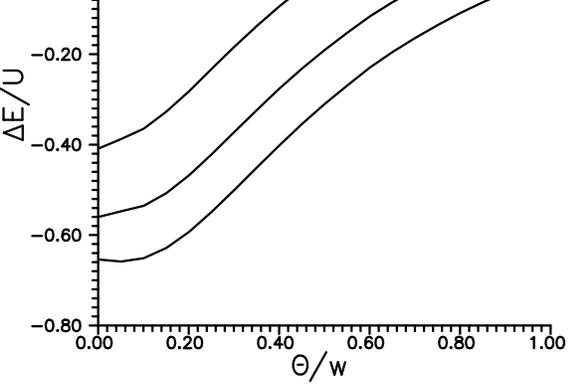
\end{document}